# Thin Film Absorbers Based on Plasmonic Phase Resonances

Yanxia Cui[1, 2], Kin Hung Fung[1], Jun Xu[1], Sailing He[2], and Nicholas X. Fang[1]*

[1] *Department of Mechanical Science and Engineering and Beckman Institute of Advanced Science and Technology, University of Illinois at Urbana-Champaign, Urbana, Illinois 61801, USA*

[2]*Centre for Optical and Electromagnetic Research, State Key Laboratory of Modern Optical Instrumentation, Zhejiang University, Hangzhou, Zhejiang, 310058, China*

*Corresponding author: nicfang@illinois.edu   phone: 1-217-265-8262

Address: 105 S Mathews Ave, MC-244, Urbana, IL 61801

We demonstrate an efficient double-layer light absorber by exciting plasmonic phase resonances. We show that the addition of grooves can cause mode splitting of the plasmonic waveguide cavity modes and all the new resonant modes exhibit large absorptivity greater than 90%. Some of the generated absorption peaks have wide-angle characteristics. Furthermore, we find that the proposed structure is fairly insensitive to the alignment error between different layers. The proposed plasmonic nano-structure designs may have exciting potential applications in thin film solar cells, thermal emitters, novel infrared detectors, and highly sensitive bio-sensors.

Recent development in thin film solar cells has gained great interest, as it could lead to economically viable solutions for solar energy conversion [1]. However, harvesting and converting sunlight into electricity becomes more challenging with use of thin active layers since absorption decreases with the film thickness. For this reason, adding metallic coatings onto a thin film solar cell has been explored for enhancing solar absorption. Although a noble metal with a flat surface reflects most of the incident light, we could dramatically enhance light absorption efficiency by structuring the metal surface. Since these metallic nanostructures support localized surface plasmons (LSP) or surface plasmon polaritons (SPP) [2, 3], they are found powerful to fold light into a very thin film, promising better absorption in these thin film solar cells [4-7].

However, challenging issues remain such as how to harvest sunlight over a broad spectrum using plasmonic structures. For example, Popov *et al.* pointed out nearly perfect absorption can be observed on a metal surface with extremely shallow grooves, which acts like a metamaterial layer [8], but this effect is only observed in a narrow wavelength range. Proposals have been made to achieve broadband light absorption over a wide solid angle, by designing a metal surface corrugated with an array of grooves of different depths [9]. However, this is difficult for current fabrication processes. Recently, metallic V-grooves (of triangular-shaped cross-sections), have also been proposed as an alternative to confine and guide light [10, 11]. However, in terms of the efficiency and spectrum bands, its absorbing performance is also limited.

In this letter, we propose a modified V-groove structure to overcome the limitation of narrow bandwidth of light absorption. We will show that by adding phase resonances, a moderate absorption peak could be split into two peaks with enhanced absorption. As a result, multiple absorption peaks could be introduced, and some of which are insensitive to the incident angle. Moreover, we study the tunability of the absorption spectra on some structural parameters.

We compare two kinds of two-dimensional metallic structures as shown in Fig. 1. The periodic T-shape groove structure in Fig. 1(a) is a modification of the original V-groove structure used in Ref. [11]. Such a structure has a spatial period of $P$, and each groove is composed of a wide groove (of width $W_1$ and height $h_1$) over a narrow groove (of width $W_2$ and height $h_2$). The other structure shown in Fig. 1(b) is our proposed structure, formed by cutting the metal horizontally at the bottleneck of the T-groove in Fig. 1(b) and drilling three additional narrow grooves. Such a structure has a narrow horizontal gap of $d$. And the additional narrow grooves are identical to the bottom part of the original T-groove and all the narrow grooves are equally separated. For simplicity, we consider all the gaps and grooves are filled with air. Our simulations show that the observed phenomena are qualitatively the same when they are filled with other dielectric media.

We consider a plane wave with TM polarization (the magnetic field is perpendicular to the *x-z* plane, defined as $H_y$) incident on the structures. Without loss of generality to the application of other noble metals, we consider the metal as silver. Permittivity parameters are obtained from Ref. [13]. The metal film is assumed to be thick enough so that there is no light transmission ($T = 0$) and the absorbtivity ($\eta$) is calculated from the reflectivity of all orders ($R$) by $\eta = 1-R$. Rigorous Coupled-wave Analysis (RCWA) [14] method is used as the simulation tool and has been verified by studying the structures in Ref. [15]. Unless otherwise mentioned, the geometrical parameters we used are $P = 1,100$ nm, $W_1 = 200$ nm, $h_1 = 300$ nm, $W_2 = 50$ nm, $h_2 = 350$ nm, and $d = 30$ nm.

The calculated absorptions at normal incidence, as a function of photon energies ($\hbar\omega$), for both metallic structures are plotted in Fig. 2 (a). It is obvious that the modified metallic structure (solid) absorbs light much better than the original one (dashed). To explain the detailed physical mechanism, we begin with a brief study of the original T-groove metallic surface. In Fig. 2(a), it is shown that this T-groove metallic surface could absorb light with an efficiency of about 65% at two frequencies (i.e., $\hbar\omega = 0.42$ and $0.81$ eV). This absorbing effect should be attributed to the plasmonic waveguide cavity modes related to the geometry of individual grooves [11, 16]. At the first absorption peak ($\hbar\omega_{01}$), each V-groove works as a quarter-wavelength cavity supporting the first order Fabry-Parot (F-P) resonance, which can be seen clearly from the distributions of field magnitude $|H_y|$ and phase $\Phi(H_y)$ in Fig. 3(a). At the second peak ($\hbar\omega_{02}$), a higher order F-P resonance arises and each groove works as a three-quarter-wavelength cavity with one field node along the vertical direction [see Fig. 3(d)]. In the left panel of Fig. 2(b), we plot the band diagram (i.e., the angular spectrum) of this T-groove surface. It is seen that the first order resonance (around 0.42 eV) has a very wide solid angle of absorption (with $\theta$ about 70º). This nearly angle-independent resonance is common for the plasmonic waveguide cavity modes. However, when the neighboring resonant cavities are coupled and interact with free photons, there is a strong dependence of the photon energy on the angle of incidence and the band becomes dispersive. This is the case for the second order resonance at about 0.81 eV. It should be noted that there is a band-folding effect in our band diagram because the period of our structure ($P$ = 1,100 nm) is comparable to the corresponding wavelength of free photons at $\hbar\omega_{02}$ ($\lambda$ = 1,530 nm). To show the coupling between free photons and the surface plasmons, we also plot the folded light-line (white) in Fig. 2. It is clear that, unlike the first order resonance, the second order resonance is disturbed by free photons so that the solid angle of absorption at $\hbar\omega_{02}$ is less than 15º.

In contrast, the absorption spectrum for our proposed structure [solid curve in Fig. 2(a)] exhibits four strong absorption peaks located at $\hbar\omega$ = 0.41, 0.47, 0.79, and 0.87 eV, labelled as $\omega_1$, $\omega_2$, $\omega_3$, and $\omega_4$, respectively. All these four peaks have theoretical absorption efficiency $\eta$ >96% and they can be grouped into two types. The two peaks at $\omega_1$ and $\omega_2$ belong to the first type, which is associated with the first order mode ($\omega_{01}$) of the original T-groove, while the other two peaks at $\omega_3$ and $\omega_4$ belong to the second type, which is associated with the second order mode ($\omega_{02}$). The band diagram of our modified structure is also shown in the right panel of Fig. 2(b) in contrast with that of the original T-groove surface. One sees that the two resonances of the first type ($\omega_1$ and $\omega_2$) are almost angle independent ($\theta$ could approach 89º), while the other two resonances of the second type ($\omega_3$ and $\omega_4$) are influenced a lot by the free-photon dispersion; this is exactly in accordance with the resonances at $\omega_{01}$ and $\omega_{02}$ for the original T-groove surface. It should be noted that the band at $\omega_4$ [dashed guide line in Fig. (2b)] is relatively flat as compared to the band at $\omega_3$. The reason for this observation will be explained later.

In order to understand the four plasmon modes of our proposed structure, we have also studied the distributions of field magnitude $|H_y|$ and phase $\Phi(H_y)$ [see Fig. 3(b), (c) ,(e), and (f)]. It should be noted that the absorption properties show a strong dependence on the total number of grooves $N$ in one period (this issue will be discussed at the end of this paper). In Fig. 1(b), we have four narrow grooves in each period (i.e., $N = 4$). We label the grooves from #1 to #4 from the left to the right as shown in Fig. 1(b) (groove #1 is separated into two halves in our defined unit cell). If we only consider the field distribution within the middle groove (#3) and the top part of the T-grooves, we see that both the magnitude and phase distributions of the first type [in Fig. 3(b) and (c)] are very close to that of the original case [in Fig. 3(a)]. This is also true for the second type [in Fig, 3(e) and (f)] versus the original case [in Fig. 3(d)]. It means that, for our proposed metallic structure, although a horizontal gap geometrically breaks the original T-groove into two parts, this gap of $d$ = 30 nm is too small to alter the field distribution within it.

Since there are additional narrow grooves that can support resonances, part of the field energy is coupled to these modes after we open the horizontal air gap. For instance, as shown in Fig. 3(b), groove #1 can have a concentrated field that is even higher than groove #3 for the peak at $\omega_1$. Such phenomenon also occurs for the peak at $\omega_4$ [see Fig. 3(f)]. This might be the reason for the relatively flat band at resonance $\omega_4$ in comparison with that at $\omega_3$ as we mentioned earlier in Fig. 2(b). At $\omega_4$, attributed to the higher energy concentration within the region of groove #1, which does not directly interact with the incident light, the whole coupling to free photons are relatively weaker and less sensitive to dispersion. However, at $\omega_3$, the electromagnetic field in the T-groove region is comparably stronger than that in other narrow grooves. This feature indicates that the whole structure would strongly interact with the incident free photon, which causes a dispersive absorption band just as that at $\omega_{02}$ for the original T-groove surface.

The additional narrow grooves, opened in the bottom layer, play a significant role of boosting absorption and overcoming the narrow bandwidth of traditional absorbers. Such pronounced effect is due to the $\pi$ phase resonance generated by these openings along the horizontal direction. In Fig. 3(b) and (c), we see that although the resonances at $\omega_1$ and $\omega_2$ are of the same type, they are different in terms of the phase information along the $x$-direction. If we consider the phase near the opening of each groove, the phase distribution at $\omega_1$ and $\omega_2$ can be represented by '- + + + -' and '+ - + - +', respectively (the sequence of signs corresponds to the field sign at the opening of grooves #1, #2, #3, #4, and the next #1 at a certain time). Here, similar to a previous work [12], we use the sign '+' to represent a phase close to $\pi$ and the sign '-' to represent a phase close to zero. At $\omega_1$ and $\omega_2$, we find that there is a phase change of $\pi$ and $2\pi$, respectively, within one period along the $x$ axis. Such phase information suggests that the resonance at $\omega_1$ and $\omega_2$ can be considered as a surface plasmon resonance with a wavenumber (or momentum) of $2\pi/P$ and $4\pi/P$, respectively. The excitation of finite parallel momentum at normal incidence is due to the lattice scattering along the $x$-axis. Such an effect is also called plasmonic phase resonances, which can give rise to rich features in the transmission spectrum, such as sharp dips [12]. Using these phase information along the $x$-direction, we can see the relationship between the resonant frequency and the number of anti-phase ($\pi$ phase) on the adjacent grooves (a typical hybridization effect). For the resonance at $\omega_2$, the field at every narrow groove is in anti-phase with its nearest neighbour, which results in a relatively high resonant frequency; this is consistent with our finding that $\omega_2 > \omega_1$ because only groove #1 is in anti-phase with its neighbours at $\omega_1$. For the second type of resonances, the phase distribution at $\omega_3$ and $\omega_4$ can be represented by '+ + + + +' and '- - + - -', respectively. In this case, there is no anti-phase adjacent groove at $\omega_3$, which is consistent with our finding that $\omega_3 < \omega_4$.

From a different point of view, we could also regard our proposal as a double-layer structure, i.e., the top nanostrip layer and the bottom nanogroove layer. Our study shows that our double-layer metallic surface works much better in comparison with the case of a single layer with only the narrow grooves which displays only one weak absorption peak ($\eta_{max} = 38\%$) at $\hbar\omega = 0.55$ eV at normal incidence (not shown in this paper). This also means that nearly three fold of improvement of absorption is achieved after adding the top nanostrip layer, in conjunction with additional peaks due to mode splitting.

An important practical issue for such a double-layer structure is that careful alignment between layers in the fabrication process may be necessary. Therefore, we study the absorption dependence on such alignment errors. We define a parameter $S_x$ as the distance between the center of the top wide air gap and the center of the bottom narrow groove (#3) along the *x*-axis. We tune $S_x$ from 0 to $P/8$ (only this range is considered because all the bottom narrow grooves are identical) and plot the absorption spectra at normal incidence in Fig. 4(a). Apparently, when $S_x$ lies in the range between 0 to 100 nm, the absorption efficiency of our system within the considered frequency range is almost invariant. When $S_x$ is larger than 100 nm, i.e., when the center of top wide air gap is far from the center of bottom narrow groove, the resonance at $\omega_2$ becomes weaker and finally disappears, while other three absorption peaks are not influenced at all. In this sense, our light absorber is quite robust and fairly insensitive to alignment error.

In Fig. 4(b), we also plot the absorption at normal incidence as the width of the top wide air gap ($W_1$) varies in the range of 100 to 500 nm. One sees that the resonances at $\omega_1$ and $\omega_2$ have only a slight blueshift when $W_1$ increases, while the resonances at $\omega_3$ and $\omega_4$ have a very small redshift for small $W_1$ and then a slight blueshift for large $W_1$. Fig. 4(b) also shows that the absorption efficiency for the resonance at $\omega_3$ becomes much weaker when $W_1$ is larger than 300 nm. Additionally, it is indicated that the spectrum is almost insensitive to the width of the top wide air gap within the range from $W_1$ = 100 nm to 300 nm.

Our study also indicates that there is a detailed balance of absorption enhancement with respect to the number of the narrow grooves in each period, as shown in Fig. 4(c). When the grooves are added gradually from 1 to 3, the number of the absorption peaks increases due to mode splitting. When $N$ equals to 4, the above-mentioned four phase states are all perfectly excited. However, further increase of groove number $N$ does not improve absorption as effectively. When $N$ is larger than 4, we see that one of the absorption peaks ($\omega_4$) becomes weaker and weaker and the other peaks ($\omega_{1-3}$) do not performs as good as that for $N = 4$ because of the phase matching property.

In conclusion, we have demonstrated an efficient light absorber by a novel hybrid grating structure inspired by V-groove metallic surfaces. By introducing a small horizontal gap and additional narrow grooves which support the excitation of plasmonic phase resonances, the overall metallic surface displays four strong absorption peaks at infrared frequencies. Our study has shown that the present structure is robust and does not require highly accurate alignment in fabrication. We have also shown that some of the generated absorption peaks exhibit wide-angle characteristic. Exploring the realm of these plasmonic nano-structure designs, we could envisage exciting applications in, e.g., thin film solar cells [1], thermal emitters [17], novel infrared detectors [18], and highly sensitive bio-sensors [19]. Moreover, such designs could inspire innovative designs of efficient absorbers at Microwave and THz frequencies [20].

This work is partially supported by the National Science Foundation (CMMI 0846771) and the National Science Foundation of China (60688401).

**Figure Captions**

Fig. 1. (Color online) Sketches of the original T-groove metallic surface (a) and our proposed light absorber (b). Insets are simplified configurations in the 2-D space. $P = 1,100$ nm, $W_1 = 200$ nm, $h_1 = 300$ nm, $W_2 = 50$ nm, $h_2 = 350$ nm, and $d = 30$ nm. Periodic boundary conditions are applied along the $x$ direction. Within a period, we label the grooves from #1 to #4 and the next period starts with label #1.

Fig. 2. (Color online) (a) Absorption spectra as a function of photon energies for the original T-groove surface (dashed line) and our proposed structure (solid line). All the peaks are labelled. (b) The band diagram for the original T-groove surface (left panel) and our proposed structure (right panel) with the white line representing the folded light line. Red lines indicate the incident angle at $\theta = 15º$ and $89º$. The dashed line in the right panel of (b) indicates the trend of the band of $\omega_4$.

Fig. 3. (Color online) Distributions of magnetic field $|H_y|$ and phase $\Phi(H_y)$ for the absorption peaks as labelled in Fig. 2(a). Insets of $\Phi(H_y)$ in (c-f) show the phase states of each resonance near the opening of the bottom narrow grooves. Signs '+' and '-' represent a phase close to $\pi$ and 0, respectively.

Fig. 4. (Color online) Absorption spectra at normal incidence (a) when the relative position of different layers along $x$ direction ($S_x$) is tuned, (b) when the width of top wide air gap ($W_1$) is tuned, and (c) when the number of narrow grooves ($N$) in each period varies. 2-D configurations for the cases of $N = 1$ to 3 are shown in the insets of (c).

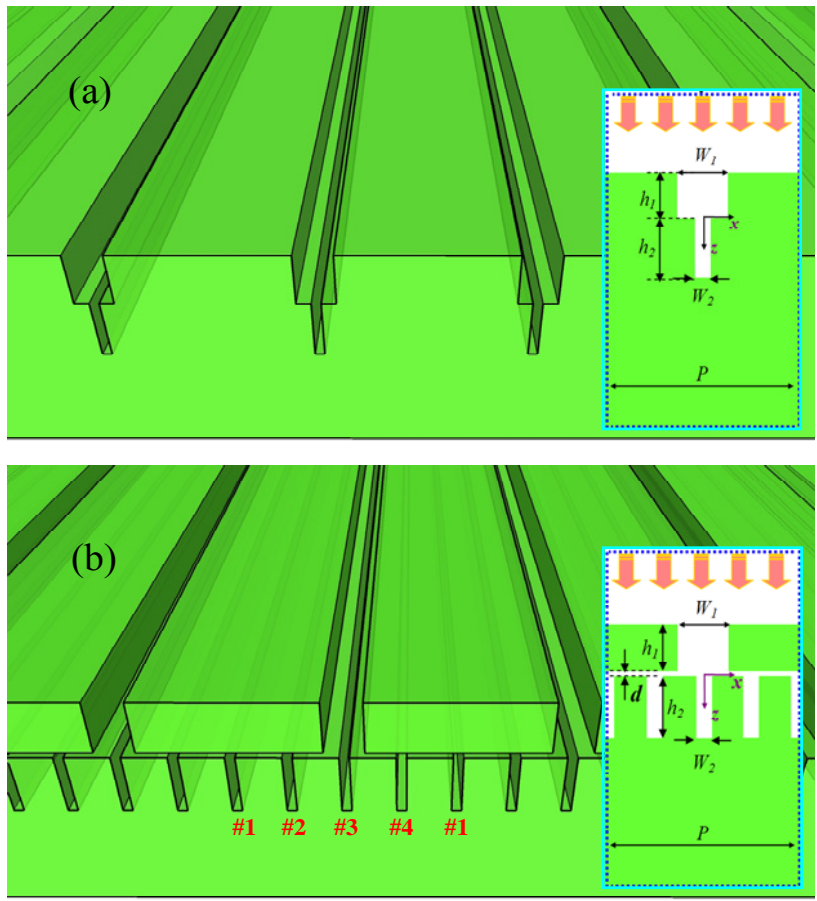

Fig. 1. Cui, Fung, Xu, He and Fang

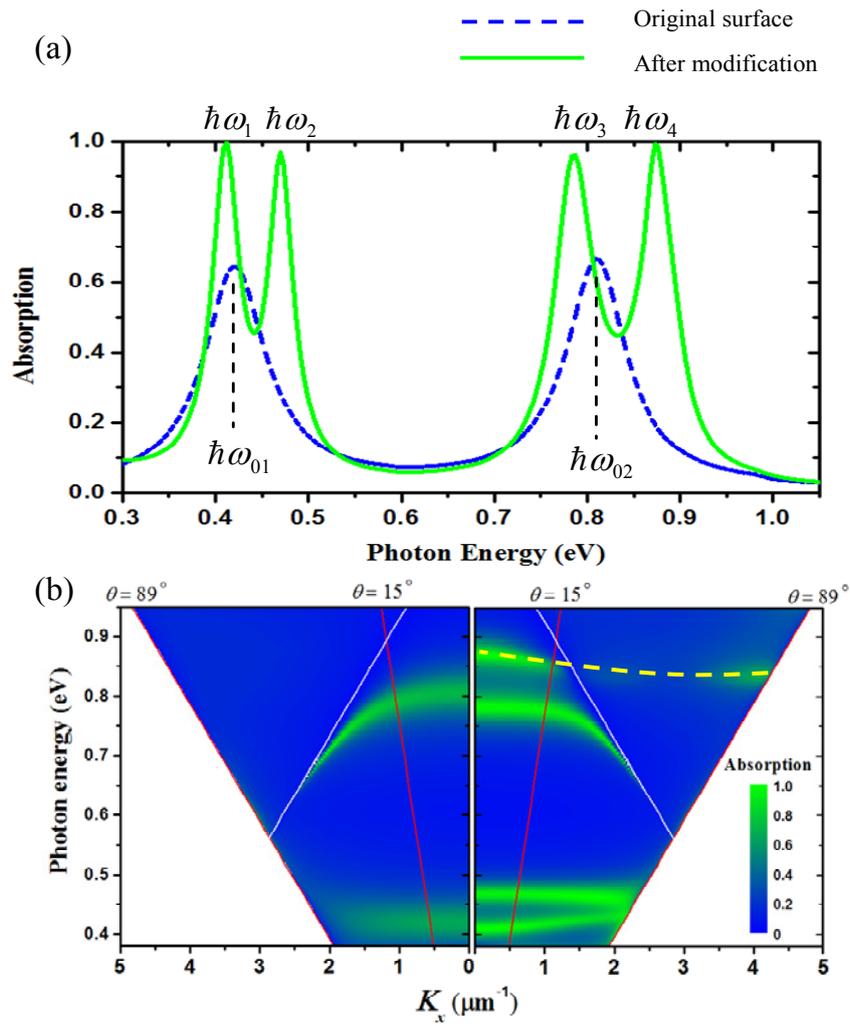

Fig. 2. Cui, Fung, Xu, He and Fang

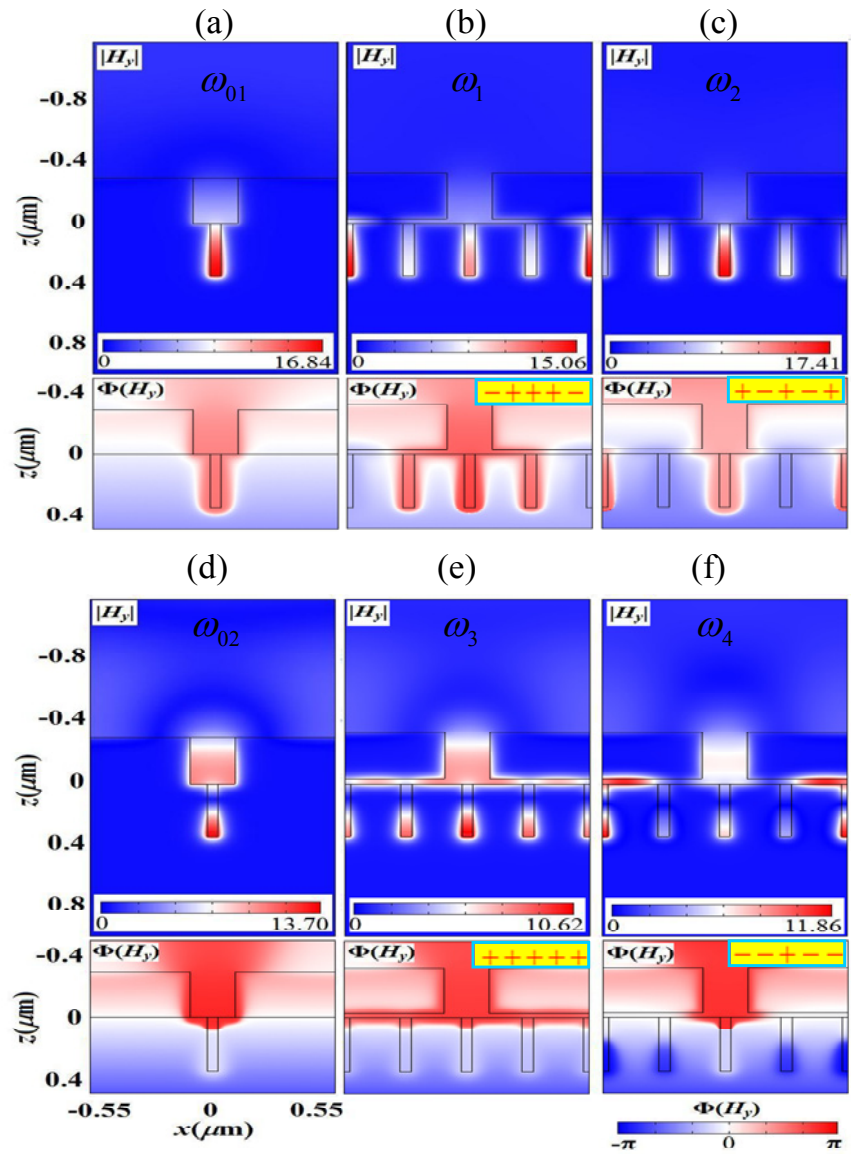

Fig. 3. Cui, Fung, Xu, He and Fang

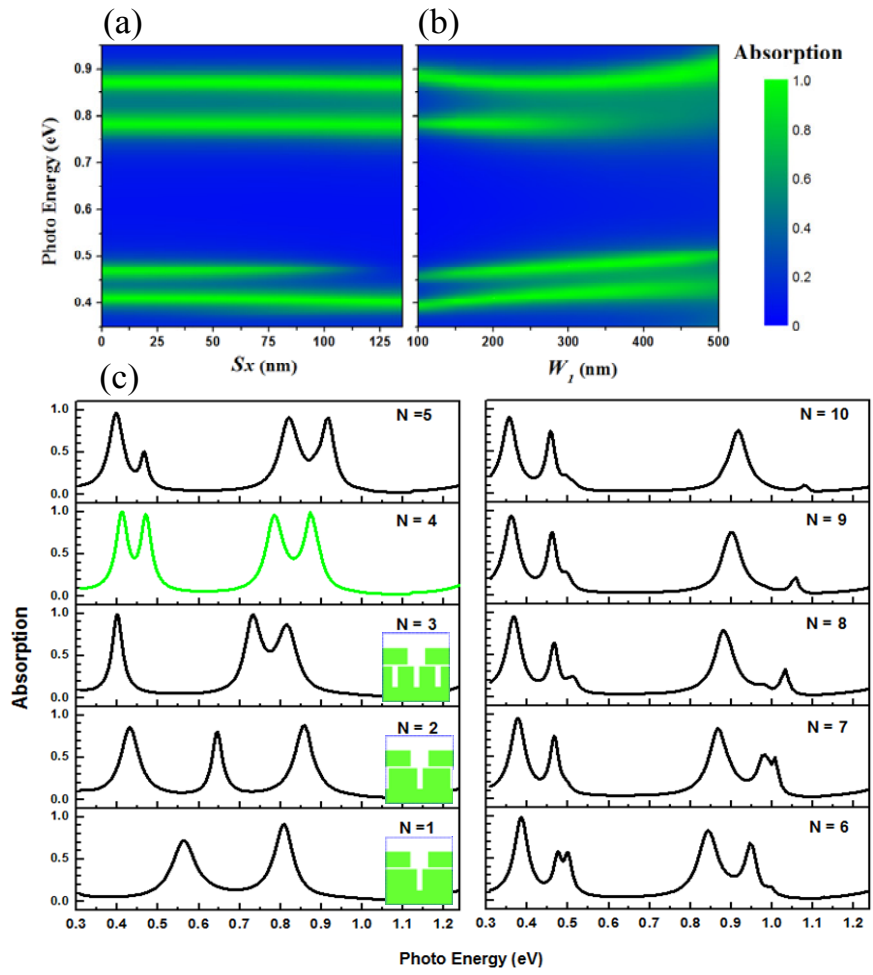

Fig. 4. Cui, Fung, Xu, He and Fang